\input harvmac

\def\bR{{\bf R}}
\def\Pf{{\rm Pf\! }}

\def\K3{{\bf K3}}
\def\journal#1&#2(#3){\unskip, \sl #1\ \bf #2 \rm(19#3) }
\def\andjournal#1&#2(#3){\sl #1~\bf #2 \rm (19#3) }

\def\hat{\widehat}

\def\tilde{\widetilde}

\def\frac#1#2{{#1\over#2}}

\def\inbar{\,\vrule height1.5ex width.4pt depth0pt}
\def\IC{\relax\hbox{$\inbar\kern-.3em{\rm C}$}}
\def\IR{\relax{\rm I\kern-.18em R}}
\def\IP{\relax{\rm I\kern-.18em P}}

%
%

%
\catcode`\@=11
\def\slash#1{\mathord{\mathpalette\c@ncel{#1}}}
\overfullrule=0pt

\def\underrel#1\over#2{\mathrel{\mathop{\kern\z@#1}\limits_{#2}}}

\catcode`\@=12


%

\def\det{{\rm det}}
\def\tr{{\rm tr}}

\def\det{{\rm det}}
\def\exp{{\rm exp}}


\def\bR{{\bf R}}


\rightline{IASSNS-HEP-00/58}
\Title{\rightline{hep-th/0008013}}
{\vbox{\centerline{A Note on Background Independence in}
\centerline{Noncommutative Gauge Theories, Matrix Model,}
\centerline{and Tachyon Condensation}{}}}
\medskip

\centerline{\it Nathan Seiberg}
\bigskip
\centerline{School of Natural Sciences}
\centerline{Institute for Advanced Study}
\centerline{Einstein Drive, Princeton, NJ 08540}

\smallskip

\vglue .3cm

\bigskip
\noindent
Using the construction of D-branes with nonzero $B$ field in the
matrix model we give a physical interpretation of the known background
independence in gauge theories on a noncommutative space.  The
background independent variables are identified as the degrees of
freedom of the underlying matrix model.  This clarifies and extends
some recent results about the end point of tachyon condensation in
D-branes with a $B$ field.  We also explain the freedom in the
description which is parametrized by a two form $\Phi$ from the points
of view of the noncommutative geometry on the worldvolume of the
branes, and of the first quantized string theory.

\Date{}

\newsec{Introduction}

D-branes with nonzero $B$ field have recently figured in different
contexts in string theory.  They were shown to arise in the matrix
model
\ref\bfss{T. Banks, W. Fischler, S.H. Shenker and  L.
Susskind, ``M Theory As A Matrix Model: A Conjecture,'' Phys. Rev.
{ \bf D55} (1997) 5112, hep-th/9610043.},
\nref\grt{O.~J.~Ganor, S.~Ramgoolam and W.~I.~Taylor,
``Branes, fluxes and duality in M(atrix)-theory,''
Nucl.\ Phys.\  {\bf B492} (1997) 191 [hep-th/9611202].}%
\nref\bss{T.~Banks, N.~Seiberg and S.~Shenker, ``Branes from
matrices,'' Nucl.\ Phys.\  {\bf B490} (1997) 91 [hep-th/9612157].}%
where an infinite collection of D0-branes can make higher Dp-branes
\refs{\bfss,\grt,\bss}.  This motivated
the discovery of noncommutative geometry in string theory
\nref\cds{A. Connes, M. Douglas and A. Schwarz, ``Noncommutative
Geometry and Matrix Theory: Compactification on
Tori,'' JHEP { \bf 02} (1998) 003, hep-th/9711162.}%
\nref\dh{M.~R.~Douglas and C.~Hull, ``D-branes and the noncommutative
torus,''  JHEP {\bf 9802} (1998) 008 [hep-th/9711165].}%
\refs{\cds,\dh}, which was later studied by various authors including
\nref\schom{V.~Schomerus, ``D-branes and deformation quantization,''
JHEP {\bf 9906}, 030 (1999) [hep-th/9903205].}%
\nref\swnonc{N.~Seiberg and E.~Witten, ``String theory and
noncommutative geometry,'' JHEP {\bf 9909} (1999) 032
[hep-th/9908142].}%
\refs{\schom,\swnonc}.  More recently, following the work of Sen
\ref\sen{A.~Sen, ``Descent Relations Among Bosonic D-branes,''
Int.J.Mod.Phys. { \bf A14} (1999) 4061-4078, hep-th/9902105;
``Universality of the tachyon potential,'' JHEP {\bf 9912} (1999) 027
[hep-th/9911116].}
\nref\senzw{A.~Sen and  B. Zwiebach, ``Tachyon condensation in string
field theory,'' JHEP { \bf 0003} (2000) 002, [hep-th/9912249].}%
\nref\mota{N.~Moeller and  W.~Taylor, ``Level truncation and the
tachyon in open bosonic string field theory,'' hep-th/0002237 .}%
\nref\DMR{K.~Dasgupta, S.~Mukhi and G.~Rajesh, ``Noncommutative
tachyons,'' JHEP {\bf 0006} (2000) 022, [hep-th/0005006].}%
\nref\chicagogro{J.~A.~Harvey, P.~Kraus, F.~Larsen and E.~J.~Martinec,
``D-branes and strings as non-commutative solitons,''
hep-th/0005031.}%
\nref\wittencond{E. Witten, ``Noncommutative Tachyons and String Field
Theory'', [hep-th/006071].}%
\nref\raszwi{L.~Rastelli and B.~Zwiebach, ``Tachyon potentials, star
products and universality,'' hep-th/0006240.}%
\nref\sochi{C.~Sochichiu, ``Noncommutative Tachyonic
Solitons. Interaction with Gauge Field,'' hep-th/0007217.}%
\nref\gms{R.~Gopakumar, S.~Minwalla and A.~Strominger, ``Symmetry
Restoration and Tachyon Condensation in Open String Theory,''
hep-th/0007226.}%
the phenomenon of tachyon condensation on D-branes was studied by
various people including \refs{\senzw-\gms}.  In particular, it became
clear that by turning on a $B$ field on D-branes, the analysis
simplifies considerably \refs{\DMR,\chicagogro,\wittencond,\gms}. The
main goal in this line of research is to learn about the ground state
of the system after tachyon condensation.  Since the open string
states are expected to disappear, there should be no dependence on the
background $B$ field.

In this note we present a unified point of view on these problems
which clarifies a number of issues.  As in the matrix model, we will
be thinking of the Dp-branes as built out of an infinite number of
D0-branes.  This perspective will make issues of background
independence manifest.  The matrix model variables are $N\times N$
matrices $X^i$.  Different Dp-branes arise as different classical
solutions $X^i_{cl}=x^i$ such that
\eqn\classsolo{[x^i,x^j]= i \theta^{ij}.}
The degrees of freedom on the Dp-branes arise by expanding the
dynamical variables around the classical solution
\eqn\dynv{X^i=x^i+\theta^{ij}\hat A_j,}
and considering $\hat A_j$ as functions of $x^i$. From this point of
view it is clear that $X^i$ are background independent and they do not
depend on $\theta$.

The subject of background independence has already appeared in the
study of gauge theories on noncommutative space.  Most of the
discussion on this subject concerns with gauge theories on a torus,
and is known in the mathematical literature as Morita equivalence.
For the case of noncommutative $\bR^p$ an explicit construction of
background independent variables was given in section 3.2 of \swnonc.
We identify them with $X^i$ of \dynv.  The relation with the matrix
model gives a physical interpretation of this mathematical
observation.

Motivated by \gms, this point of view also sheds light on the recently
studied tachyon condensation phenomenon on Dp-branes.  Here it is more
convenient to use the IKKT matrix model
\ref\ikkt{N. Ishibashi,
H. Kawai, Y. Kitazawa and A. Tsuchiya, ``A Large-N Reduced Model as
Superstring,'' Nucl. Phys. {\bf B498} (1997) 467, hep-th/9612115.},
which is based on D-1-branes. Their degrees of freedom $X^i$ provide a
background independent description of the Dp-branes. On the Dp-branes
$[X^i,X^j]\not=0$, but in the ground state without the Dp-branes
$[X^i,X^j]=0$. For the generic solution the $U(N)$ symmetry ($N\to
\infty$) of the matrix model is spontaneously broken to $U(1)^N$.  For
the special solutions with $X^i=c^i$, which are proportional to the
unit matrix the underlying $U(N)$ symmetry is unbroken. As is standard
in the IKKT model, we interpret the eigenvalues of $X^i$ as spacetime
points and identify the $U(\infty)$ symmetry of the theory with the
gauge symmetry on the Dp-branes.

\nref\piosch{B.~Pioline and A.~Schwarz,
``Morita equivalence and T-duality (or B versus Theta),''
JHEP {\bf 9908} (1999) 021, [hep-th/9908019].}%

We will also address a related topic associated with the freedom
in the description of the theory parametrized by a choice of a
two form $\Phi$ \refs{\piosch,\swnonc}.  For every background
characterized by $B$, the closed string metric $g$, and the
string coupling constant $g_s$ we have a continuum of
descriptions labeled by a choice of $\Phi$.  The open string
metric $G$, the noncommutativity $\theta$ and the open string
coupling $G_s$ are determined by \eqn\Gtheta{\eqalign{ &{1 \over
G+\Phi}+\theta  = {1 \over g+ B}\cr &G_s=g_s\left({\det(G+\Phi)
\over \det(g+ B)}\right)^{1\over 2}}} (we have set
$2\pi\alpha'=1$).  What is the geometric meaning of $\Phi$?   We
argue that on a noncommutative space the choice of $\Phi$ enters
through the commutator of two derivatives:
\eqn\basiccomm{\eqalign{ &[x^i,x^j]=i\theta^{ij}\cr
&[\partial_i,x^j]=\delta^j_i\cr
&[\partial_i,\partial_j]=-i\Phi_{ij}}.} Using these relations and
the covariant derivative $D_i=\partial_i -i\hat A_i$ we
immediately get \eqn\fhatcomm{[D_i,D_j]=-i(\hat
F_{ij}+\Phi_{ij}),} which suggests that if the action is written
in terms of $D_i$, the dependence on $\hat F$ and $\Phi$ should
be only through the combination $\hat F+\Phi$.  Indeed, it was
shown in \swnonc\ that the Dirac-Born-Infeld action depends only
on this combination, and it was suggested that this could also be
the case in higher orders.

From the string worldsheet point of view the freedom in $\Phi$ was
interpreted in \swnonc\ as a choice of regularization which is related
to field redefinition in spacetime.  Following
\ref\andbornd{O.~Andreev and H.~Dorn,
``On open string sigma-model and noncommutative gauge fields,''
Phys.\ Lett.\  {\bf B476} (2000) 402 [hep-th/9912070].}
we will present a calculation of the S-matrix which makes the
assertion about the dependence on $\hat F+\Phi$ manifest.

In the rest of this note we elaborate on these points.  In section 2
we review the construction of Dp-branes in the matrix model and the
analysis of the small fluctuations around the classical solution.  In
section 3 we discuss the relation between the underlying $U(N)$
symmetry of the matrix model and the noncommutative gauge symmetry on
the Dp-branes.  In section 4 we use this physical picture to clarify
some issues of background independence, and in section 5 we comment on
some aspects of tachyon condensation. In section 6 we discuss various
issues associated with $\Phi$.

\newsec{Dp-branes in Matrix Model}

We consider the matrix model in flat $\bR^{11}$ with transverse metric
$g_{IJ}$ ($I,J=1,...,9$).  We assume for simplicity that the metric
$g_{IJ}$ is block diagonal; i.e.\ it vanishes for $I=1,...,p$,
$J=p+1,...,9$. The dynamical variables $X^I$ are $N\times N$ hermitian
matrices.  The potential of the matrix model
\eqn\potexp{-g_{IK}g_{JL}\Tr [X^I,X^J] [X^K,X^L]}
determines the time independent equation of motion
\eqn\claeom{g_{JL}[X^J, [X^K,X^L]]=0.}
A simple set of a classical solutions are $X^I=x^I$ such that
$[x^I,x^J]$ are proportional to the unit matrix in the $N\times N$
dimensional space (such solutions exist only for infinite $N$).  Up to
translations these can be taken to be $X^i=x^i$ for $i=1,...,p$ ($p$
is even) and $X^{a+p}=0$ for $a=1,...,9-p$ with
\eqn\classsol{[x^i,x^j]= i \theta^{ij}.}
The rank of $\theta$ is $p$ (if it is smaller, it means that
effectively $p$ is smaller) and we define the $p\times p$ matrix
\eqn\invtheta{B=\theta^{-1}.}

We now expand the dynamical variables around the classical
solution. Such an expansion was first performed to leading order in
\bss, and was later extended to all orders by many authors including
\nref\li{M. Li, ``Strings from IIB Matrices,''
hep-th/9612222, Nucl. Phys. {\bf B499} (1997) 149.}%
\nref\aiikkt{H.~Aoki, N.~Ishibashi, S.~Iso, H.~Kawai, Y.~Kitazawa and
T.~Tada, ``Noncommutative Yang-Mills in IIB matrix model,''
Nucl.\ Phys.\  {\bf B565} (2000) 176 [hep-th/9908141].}%
\nref\ishibashi{N.~Ishibashi, ``A relation between commutative and
noncommutative descriptions of  D-branes,'' hep-th/9909176.}%
\nref\iikk{N. Ishibashi, S. Iso, H. Kawai and Y. Kitazawa,
``Wilson Loops in Noncommutative Yang-Mills,'' hep-th/9910004.}%
\nref\barsminic{I. Bars and  D. Minic, ``Noncommutative geometry on a
discrete periodic lattice and gauge theory,'' hep-th/9910091.}%
\nref\amns{J. Ambjorn , Y. Makeenko, J. Nishimura and R.Szabo,
``Finite $N$ matrix models of noncommutative gauge theory, ''
hep-th/9911041,  JHEP { \bf 11} 029 (1999); ``Nonperturbative dynamics
of noncommutative gauge theory,''  hep-th/0002158;
J.~Ambjorn, K.~N.~Anagnostopoulos, W.~Bietenholz, T.~Hotta and
J.~Nishimura,
``Large N dynamics of dimensionally reduced 4D SU(N) super Yang-Mills
theory,'' JHEP {\bf 0007} (2000) 013
[hep-th/0003208];
J.~Ambjorn, Y.~M.~Makeenko, J.~Nishimura and R.~J.~Szabo,
``Lattice gauge fields and discrete noncommutative Yang-Mills theory,''
JHEP {\bf 0005} (2000) 023
[hep-th/0004147].}%
\nref\agwa{L.~Alvarez-Gaume and S.~R.~Wadia, ``Gauge theory on a
quantum phase space,'' hep-th/0006219.}%
\nref\fatoll{A.~H.~Fatollahi,
``Gauge symmetry as symmetry of matrix coordinates,''
hep-th/0007023.}%
\refs{\li-\fatoll}.  Here we review the results.  We change notation to
\eqn\expacl{\eqalign{
&C_i=B_{ij}X^j=B_{ij}x^j + \hat A_i \qquad i=1,...,p\cr
&\phi^a={1 \over 2\pi \alpha'} X^{a+p}\qquad\qquad\qquad
a=1,...,9-p,}}
and calculate
\eqn\comcal{\eqalign{
&[C_i,C_j]=-iB_{ij} + B_{ik}[x^k, \hat A_i] -B_{jk}[x^k, \hat A_j]+
[\hat A_i, \hat A_j]\cr
&[C_i,\phi^a]=B_{ij}[x^j,\phi^a] + [\hat A_i, \phi^a].}}
Then the potential of the matrix model becomes
\eqn\potexpbb{\eqalign{
&-g_{IK}g_{JL}\Tr [X^I,X^J] [X^K,X^L]= (2\pi\alpha')^4
G^{ik}G^{jl}\Tr \left(\hat
F_{ij} -B_{ij}\right)\left(\hat F_{kl} -B_{kl}\right) - \cr
&\qquad \quad 2 (2\pi\alpha')^4G^{ij}g_{ab}\Tr \left(B_{ik}[x^k,\phi^a]
+ [\hat A_i, \phi^a]\right)\left(B_{jl}[x^l,\phi^b] + [\hat A_j,
\phi^b]\right) - \cr
&\qquad \quad  (2\pi\alpha')^4 g_{ac}g_{bd}\Tr [\phi^a,\phi^b]
[\phi^c,\phi^d],}}
where $\Tr$ is a trace over the $N$ dimensional matrices and
\eqn\biggd{\eqalign{
&G^{ij}= - (2\pi\alpha')^{-2}\theta^{ik}g_{kl}\theta^{lj}\cr
&\hat F_{ij}= -iB_{ik}[x^k,\hat A_j]+iB_{jk}[x^k,\hat A_i]-i[\hat
A_i,\hat A_j].}}

If $x^i$ generate the entire set of $N\times N$ matrices, every
$N\times N$ matrix $M$ can be expressed as a function of $x^i$.  We
thus generate the space of the brane as parametrized by $x^i$.  Since
the $x^i$ do not commute, the brane is noncommutative.  A convenient
ordering of $x^i$ in $M(x^i)$ is Weyl-ordering\foot{Weyl-ordering can
be defined in terms of a power series in $x^i$, with each monomial
averaged over the ordering of the factors.  Alternatively, it is
defined in terms of a Fourier transform with $e^{ip_i x^i}$ ordered as
a power series in $p_ix^i$.}.  If $M_1$ and $M_2$ are ordered
properly, the ordered product is the star product $M_1*M_2$ with the
noncommutativity parameters $\theta$.  If $x^i$ do not generate the
whole set of $N\times N$ matrices, every $N\times N$ matrix $M$ can be
expressed as a $K\times K$ matrix whose entries are functions of
$x^i$.  For simplicity we will consider the case of finite $K$.

It is easy to see that for every $K\times K$ matrix of functions
$M(x^i)$
\eqn\derdef{\eqalign{
&\partial_i M=-iB_{ik}[x^k,M] \cr
&D_i M= \partial_i M - i[\hat A_i,M]= -i[C_i,M] ,}}
where $x^i$ is proportional to the unit matrix in the $K$ dimensional
space.  In \derdef\ all the products are matrix products and $*$
products of their elements.

Using this way of representing the matrices, $\hat A_i$ and $\phi^a$
can be regarded as $K\times K$ matrices which are functions of
$x^i$.  The potential $V$ \potexp\ can now be written as
\eqn\potexpa{\eqalign{
&-g_{IK}g_{JL}\Tr [X^I,X^J] [X^K,X^L]=  \cr
&\qquad(2\pi\alpha')^4\int {d^px\over
(2\pi)^{p\over 2} } \Pf B
G^{ik}G^{jl}\tr\left(\hat F_{ij} -B_{ij}\right)\left(\hat F_{kl}
-B_{kl}\right) + \cr
&\qquad\qquad 2G^{ij}g_{ab}\tr D_i\phi^a D_j\phi^b
- g_{ac}g_{bd}\tr [\phi^a,\phi^b][\phi^c,\phi^d].}}
Here $\Tr$ is a trace over the $N\times N$ matrices and $\tr $ is a
trace over the $K\times K$ matrices.  The measure of integration is
determined as in the relation between a trace over the Hilbert space
and the integral over the classical phase space
\eqn\phassp{\Tr = \int {d^px\over
(2\pi)^{p\over 2} } {1 \over \Pf \theta} \tr= \int {d^px\over
(2\pi)^{p\over 2} } \Pf B \tr.}

It is straightforward to add the kinetic term of $X^I$ and the
fermions and to write them as an integral over the brane.  The whole
action becomes then the minimal coupling in noncommutative rank $K$
super-Yang-Mills theory.

We would like to make a number of comments:
\item{1.}  Using \phassp\ on the unit operator we deduce that ${N\over
V}={\Pf B K\over(2\pi)^{p\over 2} } $, where $V$ is the volume of the
brane.  Therefore, we can interpret $B$ as the background $B$ field on
the Dp-brane and $N$ as the induced D0-brane charge on the Dp-brane.
We learn that these Dp-branes necessarily have $B$ field on them.
\item{2.}  Restoring factors of $\alpha'$ in \Gtheta\ we get
\eqn\Gthetar{\eqalign{
&{1 \over G+2\pi \alpha' \Phi} +{\theta \over 2\pi
\alpha'} = {1 \over g+ 2\pi \alpha' B}\cr
&G_s=g_s\left({\det(G+2\pi\alpha'\Phi) \over \det(g+2\pi\alpha'
B)}\right)^{1\over 2}.}}
A natural choice of $\Phi$, which was identified in \swnonc\ is
\eqn\conchoice{\eqalign{
&\Phi=-B\cr
&\theta={1 \over B}\cr
&G=-(2\pi\alpha')^2B{1\over g} B\cr
&G_s=g_s\det(2\pi\alpha'Bg^{-1})^{1\over 2}.}}
These expressions look like the expressions in the zero slope limit,
but they are exact for all values of $\alpha'$.  The form of the
metric $G$ in \biggd\ and the fact that the action is proportional to
$(\hat F -B)^2$ show that the matrix model naturally leads to this
choice of $\Phi$.  In section 6 we will explain why the matrix model
leads to this choice.
\item{3.}  The matrix model is based on a certain scaling limit of
string theory
\nref\senli{A.~Sen, ``D0-branes on ${\bf T}^n$ and Matrix Theory,''
Adv. Theor. Math. Phys. {\bf 2} (1998) 51, hep-th/9709220.}%
\nref\seibergdlcq{N.~Seiberg, ``Why is the Matrix Model Correct?''
Phys. Rev. Lett. {\bf 79} (1997) 3577, hep-th/9710009.}%
\refs{\senli,\seibergdlcq}, which is essentially a zero slope limit,
in which the Lagrangian simplifies and is quadratic in the field
strength.  However, for the manipulations in this section this limit
is not essential.  We could have repeated the analysis replacing the
Lagrangian with the full Lagrangian of $X$ in string theory, ending
with a more complicated answer in terms of the same dynamical fields
on the Dp-brane.
\item{4.}  We can repeat the calculation for the IKKT model which is
based on D-1-branes \ikkt. Then the entire Lagrangian is given by
\potexp\ (plus fermion terms), and there is no need to include a
kinetic term.  Unlike the BFSS model, here we are not aware of a zero
slope limit which justifies the use of the minimal
Lagrangian. However, as in the case of D0-branes, the inclusion of
higher order terms do not affect our main conclusion.

\newsec{The Gauge Symmetry}

The $U(N)$ gauge symmetry of the matrix model acts on the dynamical
variables as $\delta X^I= i[\lambda, X^I]$.  If we do not want it to
act on the background $x^i$, it must act as a noncommutative gauge
transformation of rank $K$ on the D-brane \refs{\bss-\fatoll}:
\eqn\gaugsa{\eqalign{
&\delta \hat A_i =iB_{ik}[\lambda, x^k] +
i[\lambda,\hat A_i]=\partial_i \lambda+  i[\lambda,\hat A_i] \cr
&\delta \phi^a=i[\lambda,\phi^a],}}
where again the products are matrix products and $*$ products of the
elements.  We conclude that the noncommutative gauge symmetry on the
Dp-brane is the underlying $U(N)$ symmetry of the matrix model.

\nref\cornalba{L.~Cornalba, ``D-brane physics and noncommutative
Yang-Mills theory,'' hep-th/9909081.}%

For sufficiently small $\theta^{ij}\hat A_j$ all the eigenvalues of
the matrix $\partial_i X^j(x)$ are positive and we can take $X^i$ as
coordinates on the brane \refs{\ishibashi,\cornalba}.  The advantage
of doing that is that $X^i$ are background independent and can be used
to describe the coordinates on branes for all values of $\theta$.
This explains the origin of the ordinary gauge fields on the branes
\refs{\ishibashi,\cornalba} by noting that
\eqn\regare{\left[-{1\over 2} B_{ij}X^j + A_i(X)\right]\partial_l X^i
= -{1\over 2} B_{lj}x^j +{1\over 2} \partial_l (x^j\hat A_j(x)).}
Here $A_i(X)$ are ordinary commutative gauge fields with ordinary
gauge symmetry which are related to the noncommutative gauge fields
through \swnonc
\eqn\hataa{\hat A_i = A_i + {1\over 2} \theta^{ab}(2A_b\partial_a A_i
+ A_a \partial_i A_b) + \CO(\theta^2)}
and the $x$'s are treated as commuting coordinates.  (The order
$\theta^2 $ corrections to \regare\ were verified by Govindan Rajesh
\ref\grajesh{G. Rajesh, unpublished.}.)
From \regare\ we readily get
\eqn\regarea{\left[B_{ij} + F_{ij}(X)\right]\partial_k
X^i\partial_lX^j= B_{kl}.}
This can be interpreted as the statement that the coordinates $x^i$
are defined in terms of $X^i$ as the coordinates in which the
commutative field strength on the brane is constant.

This discussion makes it clear that while the noncommutative
description is always valid, the commutative description with the
coordinates $X^i$ cannot be used when $\partial_iX^j(x)$ is not
positive definite.  This is the case when the fluctuations in $X$
around the classical solution $x$ are large or when $\hat F$ is
sufficiently large.  This shows that the change of variables from
$\hat A$ and $A$ has only a finite radius of convergence.  More
precisely, we see from \regarea\ that when $\partial_iX^j(x)$ has a
zero eigenvalue, $F$ must diverge.  Similarly, when $B+F$ has a zero
eigenvalue $\partial_iX^j(x)$ must diverge.  A first sign of such a
pole has already appeared in \swnonc, where it was found that the
expression for $F$ as a function of $\hat F$ has a pole for constant
$\hat F$ when $\hat F- 1/\theta$ has a zero eigenvalue, and that $\hat
F$ diverges when $F+B$ has a zero eigenvalue.  Here we see the same
fact for any $F$ which is not necessarily constant.

\newsec{Background Independence}

Section 3.2 of \swnonc\ presents an explicit version of Morita
equivalence for the special case of noncommutative $\bR^p$ and
discusses it as background independence.  In background independence
we mean the following.  We hold the closed string metric $g$ and the
closed string coupling $g_s$ fixed and vary the background $B$ on the
D-branes.  This has the effect of changing the noncommutativity
$\theta$, as well as the open string metric $G$ and the open string
coupling $G_s$.

This background independence should not be confused with the choice of
$\Phi$ in \Gtheta.  The later keeps the background fixed and changes
the language used to describe it.

Let us first review and extend the discussion in \swnonc\ for the
special case of $\theta$ of rank $p$.  We use the coordinates $x^i$
satisfying $[x^i,x^j]=i\theta^{ij}$ to form the objects
\eqn\CXdef{\eqalign{
&C_i=\theta^{-1}_{ij}x^i+\hat A_i\cr
&X^i= \theta^{ij}C_j= x^i + \theta^{ij}\hat A_j,}}
which satisfy
\eqn\CCcom{\eqalign{
&-i[C_i,C_j]=\hat F_{ij}-\theta^{-1}_{ij}\cr
&i[X^i,X^j]=\theta^{ik} \hat F_{kl}\theta^{lj}-\theta^{ij}.}}
The main point is that the expression for $\hat F$ does not involve
explicit derivatives with respect to $x$.

We define
\eqn\partialpd{\partial_i'=\partial_i+i\theta_{ij}^{-1}x^j,}
and write the covariant derivative as
\eqn\covder{D_i=\partial_i'-iC_i(x).}
Using
\eqn\parprp{[\partial_i',x^j]=0}
we find that for every function of $x$, $M(x)$
\eqn\covderm{[D_i, M(x)]=-i[C_i(x),M(x)].}
Therefore, using \CCcom\ and \covderm, the entire Lagrangian can be
expressed in terms of $C(x)$ and perhaps other fields $M(x)$, without
explicit derivatives with respect to $x$.

The minimal $\hat F^2$ Lagrangian can be replaced by $(\hat F -
\theta^{-1})^2$, which differs from it by a constant and a total
derivative.  More generally, in string theory the Dirac-Born-Infeld
action was shown in \swnonc\ to depend only on the combination $\hat F
+\Phi$, which becomes with the natural choice \conchoice\ $\hat F-
\theta^{-1}$.

The most general Lagrangian in which $\hat F$ and $\theta$ appear only
in the combination $\hat F - \theta^{-1}$ can be written in terms of
$C$ and perhaps other fields $M(x)$ without explicit $\theta$
dependence
\eqn\lagforg{{1 \over G_s} \int d^p x \sqrt G \tr {\cal L}
(G,C(x),M(x)).}
$\theta$ enters only in the commutation relations of $x^i$.

Now we must make sure that as we vary the background we hold $g$ and
$g_s$ fixed rather than $G$ and $G_s$.  In other words, when the
Lagrangian is expressed in terms of $g$, $g_s$ and $\theta$, all the
$\theta$ dependence is in the choice of the commutation relations of
$x^i$.  This is not the case when the Lagrangian is expressed, as in
\lagforg, in terms of the metric $G$.  But because of the simple form
of \conchoice\ this can easily be fixed by expressing the Lagrangian
in terms of the variables $X^i$ and the metric $g_{ij}$.  For example,
the quartic terms can be written as
$G^{ii'}G^{jj'}[C_i,C_j][C_{i'},C_{j'}] \sim
g_{ii'}g_{jj'}[X^i,X^j][X^{i'},X^{j'}]$.  We end up with a Lagrangian
of the form
\eqn\lagforgc{{1 \over G_s} \int d^p x \sqrt G \tr {\cal L}
(g_{ij},X^i(x),M(x))
\sim {1 \over g_s } \int d^p x  {1 \over \Pf \theta} \tr {\cal L}
(g_{ij},X^i(x),M(x)).}
Note that the measure in the last expression does not have a factor of
$\sqrt g$; instead it has a factor of $ {1 \over \Pf \theta}$ which
transforms the same way under linear transformations of the $x$'s.  We
also note that the metric $g$ is used to contract the indices of the
fields $X$, while the metric $G$ and the noncommutativity
$\theta$ transform with the coordinates $x$.

From \lagforgc\ it is clear that changes in $\theta$ can be
compensated by changes in $x$ without changes in $X$, the metric $g$
and the closed string coupling $g_s$.  This proves background
independence.

The construction of the D-branes in the matrix model gives a simple
physical interpretation of this result. The matrix model does not
enjoy full background independence.  As a formulation of the theory in
the lightcone frame, it is invariant only under a few changes of the
background.  But since D-branes are solutions of the stationary
equation of motion of the model \claeom, the latter is invariant under
changes of the background, which affect only their details.  An example
of such a change is a change of the $B$ field on the D-brane, holding
the metric $g$ fixed.  Therefore, it is not surprising that the matrix
model leads to a formulation of D-branes which is invariant under
changes of $\theta$ for fixed $g$.

It is now obvious why the theory should be expressed in terms of
$X^i$.  These are the original dynamical variables of the matrix model
which are manifestly background independent.  Different values of
$\theta$ correspond to different classical solutions for the same
degrees of freedom with the same Lagrangian.  The matrix model also
makes it clear that the indices of the coordinates on the branes $x^i$
are the same as in the open string metric $G_{ij}$ and in the
noncommutativity $\theta^{ij}$, and that they are distinct from the
indices of the spacetime coordinates $X^i$ which are the same as those
in the metric $g_{ij}$.  Finally, we note that the measure in
\lagforgc\ is the same as in \phassp, or in any generalization of it
including higher order terms which are higher than quartic in $X$.

\newsec{Tachyon Condensation}

The work of Sen \sen\ and his followers on tachyon condensation in
open string theory (space filling D-branes) was simplified and
extended by turning on a nonzero $B$ field
\refs{\DMR,\chicagogro,\wittencond, \gms}.  It was convincingly argued
that the tachyon rolls to the closed string vacuum and the solitons in
this vacuum were identified with D-branes.

Our discussion based on the matrix model cannot be directly applied in
this case for several reasons.  First, this problem is not
supersymmetric, and correspondingly, the theory on the space filling
D-branes has tachyons $T$, which are not present in the matrix model.
Second, to fully use the simplification due to the $B$ field one
should turn on $B$ of maximal rank including along the time direction
(now $p$ includes the time direction).  In this case we cannot use the
zero slope limit which simplifies the effective Lagrangian and makes
it quadratic in the field strength.

\nref\gmso{R.~Gopakumar, S.~Minwalla and A.~Strominger,
``Noncommutative solitons,'' JHEP {\bf 0005} (2000) 020
[hep-th/0003160].}%
However, several lessons can still be drawn as above. (A connection
between the matrix model and this problem was anticipated in
\refs{\chicagogro,\gmso}.)  The effective Lagrangian on a single
D-brane is of the general form (we set $2\pi\alpha'=1$ and use
Euclidean signature)
\eqn\efflag{{1 \over G_s} \int d^{p}x \left[
V(T)\sqrt{\det(G_{ij}+\hat F_{ij} +\Phi_{ij})} + \sqrt G f(T)
G^{ij}D_i T D_j T + ...\right].}
We use the convenient choice
$\Phi=-B$ \conchoice, and write it in terms of our variables
$C_i=B_{ij}x^j+\hat A_i$ and $X^i=\theta^{ij}C_j$
\ref\tsdbi{A.~A.~Tseytlin, ``On non-abelian generalisation of the
Born-Infeld action in string theory,'' Nucl.\ Phys.\  {\bf B501}
(1997) 41 [hep-th/9701125].}
\eqn\efflagc{\eqalign{
&{1 \over G_s} \int d^{p}x \left[ V(T)\sqrt{\det(G_{ij} +
i[C_i,C_j] )}  - \sqrt G f(T) G^{ij}[C_i, T] [C_j,T] +
...\right]= \cr
&{\det B \over G_s} \int d^{p}x \left[ V(T)
\sqrt{\det(g^{ij} + i [X^i,X^j] )} - {1 \over \sqrt g} f(T)
g_{ij}[X^i, T] [X^j,T] + ...\right]= \cr
&{(2\pi)^{p\over 2} \over g_s}  \Tr \left[ V(T)
\sqrt{\det(\delta_i^{j} + i g_{ik}[X^k,X^j] )} - f(T) g_{ij}[X^i, T]
[X^j,T] + ... \right].}}
In the last expression we represented $X^i$ and $T$ as $N\times N$
matrices in the $N \to \infty$ limit.  We see that, as expected, when
the action is expressed in terms of the closed strings metric $g$, the
string coupling $g_s$, and the dynamical variables $X^i$, rather than
in terms of $G$ and $G_s$, there is no dependence on $\theta$ or $B$
(this is not the case when the action is expressed in terms of $C_i$).
This is a general result which persists even in the higher order
terms, as it follows simply from the general properties of
$X^i=x^i+\theta^{ij}\hat A_i$ in noncommutative gauge theories, or
from our underlying matrix model interpretation.

Assuming that $V(T)$ has a unique minimum $T_c$, at the vacuum $T=T_c$
is proportional to the unit matrix.  The vacuum is commutative and is
characterized by $X^i=X^i_c$ satisfying
\eqn\closco{[X^i_c,X^j_c]=0.}
Depending on the eigenvalues of $X^i_c$ the full $U(N)$ symmetry is
broken to $U(N_1)\times U(N_2) ...$. In one extreme case it is
$U(1)^N$, and in the other, when all $X^i_c$ are proportional to the
unit matrix, $U(N)$ is unbroken.

In a recent paper \gms\ it was suggested that the classical solution
in the ground state has $X^i_c=0$.  We add to this the other solutions
satisfying \closco.  These include other $U(N)$ invariant solutions,
where $X^i_c$ are all proportional to the unit matrix and other
solutions which break the $U(N)$ symmetry.  Motivated by our
discussion in the previous sections, we also add the interpretation of
the $U(N)$ symmetry as the symmetry of an underlying matrix model.
Correspondingly, we interpret the eigenvalues of the classical
solutions $X^i_c$ as $N$ spacetime points.  Section 3 makes it clear
that the $U(N)$ symmetry is not bigger than the rank $K$
noncommutative gauge symmetry on the D-branes.  In fact, these two
symmetries are the same symmetry.  Finally, the discussion in section
4 of background independence makes the fact that the answers are
independent of $\theta$ manifest.

\newsec{Comments on $\Phi$}

As we mentioned in the introduction, we interpret $\Phi$ as appearing
in the defining commutation relations \basiccomm
\eqn\basiccomms{\eqalign{ &[x^i,x^j]=i\theta^{ij}\cr
&[\partial_i,x^j]=\delta^j_i\cr
&[\partial_i,\partial_j]=-i\Phi_{ij}.}}
We assume, for simplicity, that $\theta$ is of maximal rank and
define, as in \partialpd \swnonc
\eqn\partialpdn{\partial_i'=\partial_i+i\theta_{ij}^{-1}x^j,}
which satisfy
\eqn\parprp{\eqalign{
&[\partial_i',x^j]=0\cr
&[\partial_i',\partial_j']=-i(\Phi_{ij}+\theta_{ij}^{-1}).\cr}}
The covariant derivatives
\eqn\covdern{D_i=\partial_i -i\hat
A_i=\partial_i'-iC_i(x)}
satisfy
\eqn\fhatcommn{[D_i,D_j]=-i(\hat F_{ij}+\Phi_{ij}).}

Three special cases are of particular importance.
\item{1.} $\theta=0$. This is the commutative theory.  Here we
cannot define $\partial'$, but on the other hand, we can define
$\tilde \partial_i=\partial_i-{i\over 2} \Phi_{ij}x^j$ such that
\eqn\tilparrel{\eqalign{
&[\tilde \partial_i,x^j]=\delta_i^j \cr
&[\tilde \partial_i,\tilde \partial_j]=0.}}
Physically, this is the familiar case of electrons in background
magnetic field $\Phi$, and $\tilde \partial$ is the ordinary
derivative. 
\item{2.} $\Phi=0$.  This is the ordinary noncommutative theory
whose derivatives commute.
\item{3.} $\Phi=-\theta^{-1}$.  In this case the derivatives
$\partial'$ commute both with the coordinates $x$ and with themselves.
The phase space generated by $x^i$ and $\partial_i$ is degenerate and
we can set $\partial_i'$ to zero. In other words, the algebra is the
same as with $x^i=i\theta^{ij}\partial_j$. If we consider a quantum
system based on this phase space, the wave functions are not functions
of all the $x'$s but only of half of them. Physically, this choice
arises in the problem of electrons constrained to be in the first
Landau level, where the wave functions depend only on half of the
coordinates.

Now it is clear why in the matrix model we automatically found the
third choice $\Phi=-\theta^{-1}$.  In that problem we started with
D0-branes and created the worldvolume of the Dp-branes.  Derivatives
on that worldvolume appeared, as in \derdef, as commutators with the
coordinates.  Therefore, the coordinates and the derivatives are not
independent and a linear combination of them $\partial'$ must be set
to zero.

\nref\ft{E.S. Fradkin, A.A. Tseytlin,
``Nonlinear Electrodynamics from Quantized Strings'',
Phys.Lett.{\bf B163} (1985) 123.}%
\nref\callan{C.G. Callan, C. Lovelace, C.R. Nappi, S.A. Yost,
``Open Strings in Background gauge Fields'',
Nucl.Phys.{ \bf B288} (1985) 525.}%

We would like to end this section about $\Phi$ by commenting about the
way it arises in the first quantized description of string theory.  It
was suggested in \swnonc\ that the freedom in $\Phi$ is associated
with a choice of regularization in the worldsheet path integral.  This
point was made more explicit in \andbornd.  Setting $2\pi\alpha'=1$
the propagator of the worldsheet fields $X$ along the boundary is
given by \refs{\ft,\callan}
\eqn\xprop{<X^i(\tau)X^j(0)>=-{1\over 2\pi}G_0^{ij}\log\tau^2+
{i\over 2} \theta_0^{ij}\epsilon(\tau),}
where
\eqn\Gthetaz{\eqalign{
&{1 \over G_0}=\left({1 \over g+ B}\right)_S\cr
&\theta_0  =\left({1 \over g+ B}\right)_A.}}
It was important in the discussion in \swnonc\ that the second
term in the propagator affects correlation functions of vertex
operators at separated points $<...>_{g,B}$ only in a prefactor
\eqn\corrfunc{<...>_{g,B}= e^{-{i\over 2}\sum_{n>m} k^n_i
\theta_0^{ij}k^m_j\epsilon(\tau_n-\tau_m)}<...>_{G_0,0},}
where $k^n$ are the spacetime momenta of the operators.  This
observation led to the conclusion that if the action is written in
terms of $G_0$, $\theta_0$ and $G_s$, the noncommutativity $\theta_0$
enters only in the star product.  

Let us examine what happens as we vary $\Phi$, $G$ and $\theta$ while
holding $g$ and $B$ fixed.  The symmetric and antisymmetric parts of
\eqn\Gthetan{{1 \over G+\Phi} +\theta  = {1 \over g+B}}
are
\eqn\Gthetazg{\eqalign{
&\left({1 \over G+\Phi}\right)_S={1 \over G_0}\cr
&\left({1 \over G+\Phi}\right)_A=-\theta+\theta_0,}}
and therefore
\eqn\corrfuncn{<...>_{G,\Phi}= e^{-{i\over 2}\sum_{n>m} k^n_i
(\theta_0-\theta)^{ij}_0k^m_j\epsilon(\tau_n-\tau_m)}<...>_{G_0,0}}
and
\eqn\corrfuncf{<...>_{g,B}= e^{-{i\over 2}\sum_{n>m} k^n_i
\theta^{ij}k^m_j\epsilon(\tau_n-\tau_m)}<...>_{G,\Phi}.}

We now add Chan-Paton factors to the external legs.  Each order of
these factors is associated with an order of the $\tau_n$'s.  To
find the S-matrix for a specific order of the Chan-Paton factors
we integrate over the $\tau_n$'s in a given order.  We find a
product of the standard phase of noncommutative theories $\exp
(-{i\over 2}\sum_{n>m} k^n_i \theta^{ij}k^m_j )$ with the
parameter $\theta$, and another factor which depends only on $G$
and $\Phi$.  The latter factor is the same as in a commutative
theory ($\theta=0$) with closed string metric $G$ and $B$ field
given by $\Phi$.  We stress that the S-matrix is universal and
does not depend on a choice of regularization.  This fact is
usually proven by going to a convenient region in momentum space
where the contribution from the boundary of the integration
region vanishes.

We now look for an effective action which reproduces this
S-matrix. Here we face the known ambiguity associated with field
redefinition. Ignoring the phase factor $\exp (-{i\over
2}\sum_{n>m} k^n_i \theta^{ij}k^m_j)$, we can use the ordinary
action which depends only on the metric $G$ and the combination
$F+\Phi$.  (Here is where we make a choice because there are also
other actions which lead to the same S-matrix.)  The phase is
then added by making the theory noncommutative and turning $F$
into $\hat F$.  We end up with an action which depends on $\Phi$
only through the combination $\hat F+\Phi$ and on $\theta$ only
through the star product.  Finally, the expression for the open
string coupling \Gtheta\ is determined by evaluating the action
for $\hat F=0$. This completes the proof of the conjectured form
of the action \swnonc.

\centerline{\bf Acknowledgements}
We thank T. Banks, R. Gopakumar, J. Harvey, S. Minwalla, G. Rajesh,
A. Strominger, and E. Witten for discussions.  This work was supported
in part by grant \#DE-FG02-90ER40542.

\listrefs

\end